\documentclass[11pt,letter]{article}
\usepackage{geometry}                
\usepackage{amssymb}
\usepackage{epstopdf}

\usepackage[sf,bf]{caption}

\usepackage{gensymb}

\usepackage{graphics,graphicx}

\newcommand{\OI}{O\,{\sc i}}

\newcommand{\CII}{C\,{\sc ii}}
\newcommand{\HII}{H\,{\sc ii}}
\newcommand{\HI}{H\,{\sc i}}
\newcommand{\SII}{S\,{\sc ii}}

\title{\textbf{Compression and ablation of the\\
 photo-irradiated cloud the Orion Bar}\footnote{Manuscript accepted for publication in Nature. Submitted 13 February; accepted 8 June 2016. \hspace{1cm}
 This is the author's submitted version. Definitive version of the manuscript was published online by Nature on 10 August 2016 (doi:10.1038/nature18957).}}

\author{Javier R. Goicoechea$^a$, J. Pety$^{b,c}$, S. Cuadrado$^a$, J. Cernicharo$^a$, 
E. Chapillon$^{b,d,e}$,\\ A. Fuente$^{f}$, M. Gerin$^{c,g}$, C. Joblin$^{h,i}$, 
N. Marcelino$^{a}$ and P. Pilleri$^{h,i}$}

\begin{document}
\maketitle

\vspace{1cm}

$^a$Grupo de Astrof\'{\i}sica Molecular. Instituto de Ciencia de Materiales de Madrid (CSIC). Calle Sor Juana Ines de la Cruz 3, E-28049 Cantoblanco, Madrid, Spain.\\

$^b$Institut de Radioastronomie Millim\'etrique, 300 rue de la Piscine, F-38406 Saint Martin d'H\`eres, France.\\

$^c$LERMA, Observatoire de Paris, CNRS UMR 8112, \'Ecole Normale Sup\'erieure, PSL research university, 24 rue Lhomond, 75231, Paris Cedex 05, France.\\

$^d$Universit\'e de Bordeaux, LAB, UMR 5804, F-33270, Floirac, France.\\

$^e$CNRS, LAB, UMR 5804, F-33270 Floirac, France.\\

$^f$Observatorio Astron\'omico Nacional. Apartado 112, 28803, Alcal\'a de Henares, Spain.\\

$^g$Sorbonne Universit\'es,  UPMC Universit\'e Paris 06, 75000, Paris, France.\\

$^h$Universit\'e de Toulouse, UPS-OMP, IRAP, 31028, Toulouse, France.\\

$^i$CNRS, IRAP, 9 Avenue du Colonel Roche, BP 44346, 31028 Toulouse, France.\\


\clearpage

\begin{abstract}

\large

\vspace{1cm}
The Orion Bar is the archetypal edge-on molecular cloud surface illuminated by strong ultraviolet radiation from nearby massive stars. Owing to the close distance to Orion (about 1,350 light-year), the effects of stellar feedback on the parental cloud can be studied in detail. Visible-light observations of the Bar\cite{Walmsley00} show that the transition between the hot ionised gas and the warm neutral atomic gas (the ionisation front) is spatially well separated from the transition from atomic to molecular gas  (the dissociation front): about 15 arcseconds or 6,200 astronomical units (one astronomical unit is the Earth-Sun distance). Static equilibrium models\cite{Tielens85,Andree-Labsch14} used to interpret previous far-infrared and radio observations of the neutral gas in the Bar\cite{Tielens93,Hogerheijde95,Young00} (typically at 10-20 arcsecond resolution) predict an inhomogeneous cloud structure consisting of dense clumps embedded in a lower density extended gas component.  Here we report one-arcsecond-resolution millimetre-wave images that allow us to resolve the molecular cloud surface. In contrast to stationary model predictions\cite{Sternberg95,LePetit06,Rollig07} there is no appreciable offset between the peak of the H$_2$ vibrational emission (delineating the H/H$_2$ transition) and the edge of the observed CO and HCO$^+$ emission. This implies that the H/H$_2$ and C$^+$/C/CO transition zones are very close. These observations reveal a fragmented ridge of high-density substructures, photoablative gas flows and instabilities at the molecular cloud surface. The results suggest that the cloud edge has been compressed by a high-pressure wave that currently moves into the molecular cloud. The images demonstrate that dynamical and nonequilibrium effects are important for the cloud evolution. 
\end{abstract}


\vspace{1cm}

          The ALMA radiotelescope allows us to resolve the atomic to molecular gas transition at the edge of the Orion molecular cloud\cite{Genzel89,Goicoechea15,ODell01,Werf13} that is directly exposed to energetic radiation from the Trapezium stars (Fig.~1). The strong ultraviolet (UV) field drives a blister “\HII~region”  (hot ionised hydrogen gas or H$^+$) that is eating its way into the parental molecular cloud. At the same time, flows of ionised gas stream away from the cloud surface at about 10~km\,s$^{-1}$ (roughly the speed of sound $c_{\rm HII}$ at $T$$\approx$10$^4$~K)\cite{Genzel89,Goicoechea15}.  The so-called photon-dominated or photo-dissociation region (PDR\cite{Hollenbach99}; see sketch in Extended~Data~Fig.~1) starts at the \HII~region/cloud boundary where only far-UV radiation penetrates the ``neutral'' cloud, i.e. stellar photons with energies below 13.6~eV that cannot ionise H atoms but do dissociate molecules  \mbox{(H$_2$ + photon $\rightarrow$ H + H)}, and ionise elements such as carbon \mbox{(C + photon $\rightarrow$ C$^+$ + electron)}.  Inside the PDR, the far-UV photon flux gradually decreases due to dust grain extinction and H$_2$ line absorption, and so does the gas and dust temperatures\cite{Hollenbach99}. These gradients produce a layered structure with different chemical composition as one moves from the cloud edge to the interior\cite{Hogerheijde95,Werf13}. The ionised nebula (the 
\HII~region) can be traced by the visible-light emission from atomic ions (such as the [\SII]\,6,731\,\AA~electronic line). The ionisation front is delineated by the [\OI]\,6,300\,\AA~line of neutral atomic oxygen\cite{Weilbacher15} (Fig.~1). Both transitions are excited by hot temperature collisions with electrons. Therefore, their intensities sharply decline as the electron abundance decreases by a factor of $\sim$10$^4$ at the H$^+$/H transition layer. In Fig.~1b, the dark cavity between the ionisation front and the HCO$^+$ emitting zone is the neutral ``atomic layer'' \mbox{($x$(H)$>$$x$(H$_2$)$\gg$$x$(H$^+$)} where $x$ is the species abundance with respect to H nuclei). This layer is very bright in mid-IR polycyclic aromatic hydrocarbon (PAH) emission, and cools via far-infrared O and C$^+$ emission lines\cite{Hollenbach99}. Although most of the electrons are provided by the ionisation of C atoms (thus $x$($e^-$)$\approx$$x$(C$^+$)$\approx$10$^{-4}$)\cite{Hollenbach99,Draine11}, the gas is mainly heated by collisions with energetic (about 1~eV) electrons photo-ejected from small grains and PAHs\cite{Tielens85,Hollenbach99}. For the strong far-UV radiation flux impinging the Bar\cite{Andree-Labsch14,Hogerheijde95}, approximately 4.4$\times$10$^4$ times the average flux in a local diffuse interstellar cloud\cite{Draine11}, a gas density $n_{\rm H}=n({\rm H})+2n({\rm H_2})$ of (4-5)$\times$10$^4$~cm$^{-3}$ in the atomic layer is consistent with the observed separation between the ionisation and dissociation fronts\cite{Andree-Labsch14,Tielens93}.  

\vspace{0.5cm}
ALMA resolves the sharp edge where the HCO$^+$ and CO emission becomes intense (Fig.~2). These layers spatially coincide with the brightest peaks of H$_2$ vibrational emission (H$_{2}^{*}$) tracing the H/H$_2$ transition (Extended~Data~Fig.~2). Therefore, the H/H$_2$ and the C$^+$/C/CO transition zones occur very close to each other. Static equilibrium models of a PDR with \mbox{$n_{\rm H}$=(4-5)$\times$10$^{4}$~cm$^{-3}$} predict, however, that the C$^+$/CO transition should occur deeper inside the molecular cloud because of the lower ionisation potential of C atoms (11.3~eV), and because CO may not self-shield from photodissociation as effectively as H$_2$\cite{Tielens93,Rollig07,Hollenbach99}.  The spatial coincidence of several H$_{2}^{*}$ and HCO$^+$ emission peaks shows that the formation of carbon molecules readily starts at the surface of the cloud (initiated by reactions of C$^+$ with H$_2$). This shifts the C$^+$/CO transition closer to the ionisation front and suggests that dynamical effects are important\cite{Bertoldi96,Storzer98}.

\vspace{0.5cm}
To zero order, the CO $J$=3-2 line intensity peak ($T_{\rm peak}^{{ \rm CO\,3-2}}$ in K) is a measure of the gas temperature $T$ in the molecular cloud ($\delta x >15''$ in 
Fig.~2c, where $\delta x$ is the distance to the ionisation front). The HCO$^+$ $J$=4-3 integrated line intensity ($W_{4-3}^{{\rm HCO^+}}$in K~km~s$^{-1}$), however, scales with the gas density $n_{\rm H}$ (see Methods and Extended Data Fig.~3). Although the 
$T_{\rm peak}^{{ \rm CO\,3-2}}$ image shows a relatively homogeneous temperature distribution, the $W_{4-3}^{{\rm HCO^+}}$ image shows small-scale structure (Fig. 2a, 2b). In particular, ALMA resolves several bright HCO$^+$ emission peaks (filamentary substructures, some akin to globulettes) surrounding the dissociation front and roughly parallel to it. These substructures are surrounded by a  lower-density gas component, with 
$n_{\rm H}$$\approx$(0.5--1.0)$\times$10$^5$~cm$^{-3}$, producing an extended (ambient) emission\cite{Tielens93,Hogerheijde95}. The HCO$^+$ substructures (with a typical width of about 2$''$$\approx$4$\times$10$^{-3}$\,pc) are located at the molecular cloud edge, and are different from the bigger (5$''$-10$''$) condensations previously seen deeper inside the molecular cloud\cite{Young00,Lis03}. 

\vspace{0.5cm}
To investigate the molecular emission stratification inside the cloud, we constructed averaged emission cuts perpendicular to the Bar. Three emission maxima are resolved in the $W_{4-3}^{{\rm HCO^+}}$ crosscuts at roughly periodic separations of $\sim$5$''$$\approx$0.01~pc (Fig.~2c). Excitation models show that the average physical conditions that reproduce the mean CO and HCO$^+$ intensities towards dissociation front (at $\delta x$$\approx$15$''$) are $T$$\approx$200-300~K and                           $n_{\rm H}$$\approx$(0.5-1.5)$\times$10$^6$~cm$^{-3}$ (see Methods and Extended Data Fig.~3). Hence, the over-dense substructures have compression factors of $\sim$5--30 with respect to the ambient gas component, and are submitted to high thermal pressures ($P/k$$=$$n_{\rm H}$$\cdot$$T$$\approx$2$\times$10$^8$~K~cm$^{-3}$).  The three periodic maxima suggest that a high-pressure compression wave exists, and is moving into the molecular cloud. This wave may be associated with an enhanced magnetic field (several hundred $\mu$Gauss; see Methods). 

\vspace{0.5cm}
In the very early stages of an \HII~region expansion upon molecular clouds, theory predicts that the ionisation and dissociation fronts are co-spatial (an $R$-type front\cite{Weilbacher15,Spitzer78}).  Soon after ($t$$<$10$^3$~yr), the expansion slows down and the dissociation front propagates ahead of the ionisation front and into the molecular cloud\cite{Draine11,Bertoldi96}. The ionisation front changes to a $D$-type front (a compressive wave travels ahead of the ionisation front and the neutral gas becomes denser than the ionised gas\cite{Draine11,Spitzer78}). For a front advancing at a speed of 0.5-1.0~km~s$^{-1}$\cite{Bertoldi96,Storzer98}, the observed separation between the ionisation and dissociation fronts in the Bar implies a crossing-time of 25,000-50,000 yr. For later times in the expansion phase, when $t$ is several times the dynamical time ($t_{\rm dyn}$) of the expanding \HII~region (the ratio of the initial \HII~region radius, so-called the Str{\"o}mgren radius, and the speed of sound $c_{\rm HII}$), the compressive wave slowly enters into the molecular cloud 
($t_{\rm dyn}$$\approx$0.2~pc per 10~km~s$^{-1}$$\approx$20,000~yr for the Bar)\cite{Hill78,Hosokawa06}. Observational evidences of such dynamical effects are scarce.

\vspace{0.5cm}
In the compressed layers suggested by ALMA (where $\delta$$x$ is between 7$''$ and 30$''$ in 
Fig.~2a), the distribution of gas densities follows a relatively narrow log-normal distribution (Fig.~2d). This is consistent with magneto-hydrodynamic simulations of non-gravitating turbulent clouds\cite{Hennebelle12,Federrath13}. When the entire observed field is analysed, the shape of the distribution is closer to a double-peaked log-normal distribution. This resembles specific simulations in which the cloud compression is induced by the expansion of the ionised gas\cite{Federrath13,Tremblin12} (and not by a strong turbulence). Searching for additional support to this scenario, we investigated the degree of turbulence and compared the different contributions to the gas pressure in the PDR (Extended Data Table~1). The inferred non-thermal (turbulent) velocity dispersion, about 1~km~s$^{-1}$, results in a moderate Mach number of $\lesssim$1 (the ratio of the turbulent velocity dispersion to the local speed of sound), i.e. only a gentle level of turbulence. The thermal pressure exerted by the \HII~region at the H$^+$/H interface\cite{Walmsley00} is several times higher than the turbulent and thermal pressures in the ambient molecular cloud. These pressure differences, together with the detection of over-dense substructures close to the cloud edge, agree with the UV radiation-driven compression scenario\cite{Tremblin12,Gorti02}.  Whether these substructures can be the seed of future star-forming clumps (e.g. by merging into massive clumps) is uncertain\cite{Hosokawa06,Elmegreen77}. At least presently, gravitational collapse is not apparent from their density distribution (no high-density power-law tail\cite{Federrath13,Tremblin12}). Indeed, their estimated masses (less than about 0.005~M$_{\odot})$ are much lower than the mass needed to make them gravitationally unstable. Even so, the increased UV-shielding produced by the ridge of high-density substructures likely contributes to protect the molecular cloud against photo-destruction for longer times.  

\vspace{0.5cm}
The ALMA images also show CO emission ripples\cite{Berne10} along the molecular cloud surface (undulations separated by less than about 5$''$$\approx$0.01~pc in Fig.~2b) indicative of instabilities at the dissociation front. Such small-scale corrugations resemble the ``thin-shell'' instability produced by the force imbalance between thermal (isotropic) and ram pressure (parallel to the flow)\cite{Garcia96}. Characterising these interface instabilities in detail would require new magneto-hydrodynamic models including: i) mesh-resolutions that are well below the  \mbox{0.1-0.01~pc} scales achieved in current simulations\cite{Tremblin12}, and ii) include neutral gas thermochemistry.

\vspace{0.5cm}
Finally, ALMA reveals fainter HCO$^+$ and CO emission in the atomic layer (HCO$^+$ globulettes and plume-like CO features at $\delta$$x$$<$15$''$, see Fig.~2a, 2b). The dense gas HCO$^+$ emission structures must have survived the passage of the dissociation front\cite{Lefloch94}, whereas the CO plumes may trace either warm CO that \textit{in situ} reforms in the atomic layer, or molecular gas that advects or photoablates\cite{Berne10} from the molecular cloud surface. In the latter case, the pressure difference between the compressed molecular layers and the lower density atomic layer would favour such a flow.   Interestingly, molecular line profiles from the plumes typically show two velocity components, one of them identical to that of gas from inside the Bar (Extended Data Fig.~4).  This kinematic association supports the presence of photoablative flows through the atomic layer, and overall agrees with the suggested role of dynamical and non--equilibrium effects in UV-irradiated clouds.



\clearpage

\begin{description}
 \item [Acknowledgements] 
 We thank the ERC for support under grant ERC-2013-Syg-610256-NANOCOSMOS. We also thank Spanish MINECO for funding support under grants CSD2009-00038 and AYA2012-32032. This work was in part supported by the French CNRS program ``Physique et Chimie du Milieu Interstellaire''. We thank P.~Schilke and D.~Lis for sharing their IRAM-PdBI observations of the H$^{13}$CN J=1-0 condensations inside the Bar, and M.~Walmsley for sharing his H$_2$ $v$=1-0 $S$(1) and \OI~1.3\,$\mu$m infrared images. ALMA is a partnership of ESO (representing its member states), NSF (USA) and NINS (Japan), together with NRC (Canada), NSC and ASIAA (Taiwan), and KASI (Republic of Korea), in cooperation with the Republic of Chile. The Joint ALMA Observatory is operated by ESO, AUI/NRAO and NAOJ. This paper makes use of observations obtained with the IRAM--30m telescope. IRAM is supported by INSU/CNRS (France), MPG (Germany), and IGN (Spain). 
 \item[Competing Interests] The authors declare that they have no
competing financial interests.
 \item[Author contributions] J.~R.~G. was the  PI of the ALMA project. He led the scientific analysis, modelling and wrote the manuscript. J.~P. and E.~C. carried out the ALMA data calibration and data reduction. S.~C. and N.~M. carried out the single-dish maps observations with the IRAM-30m telescope. All authors participated in the discussion of results, determination of the conclusions and revision of the manuscript.
  \item[Author information] The authors declare no competing financial interests. Correspondence and requests for materials should be addressed to  jr.goicoechea@icmm.csic.es. This paper makes use of the following ALMA data: ADS/JAO.ALMA$\#$2012.1.00352.S. 
\end{description}

\begin{figure}
\includegraphics[width=1\textwidth]{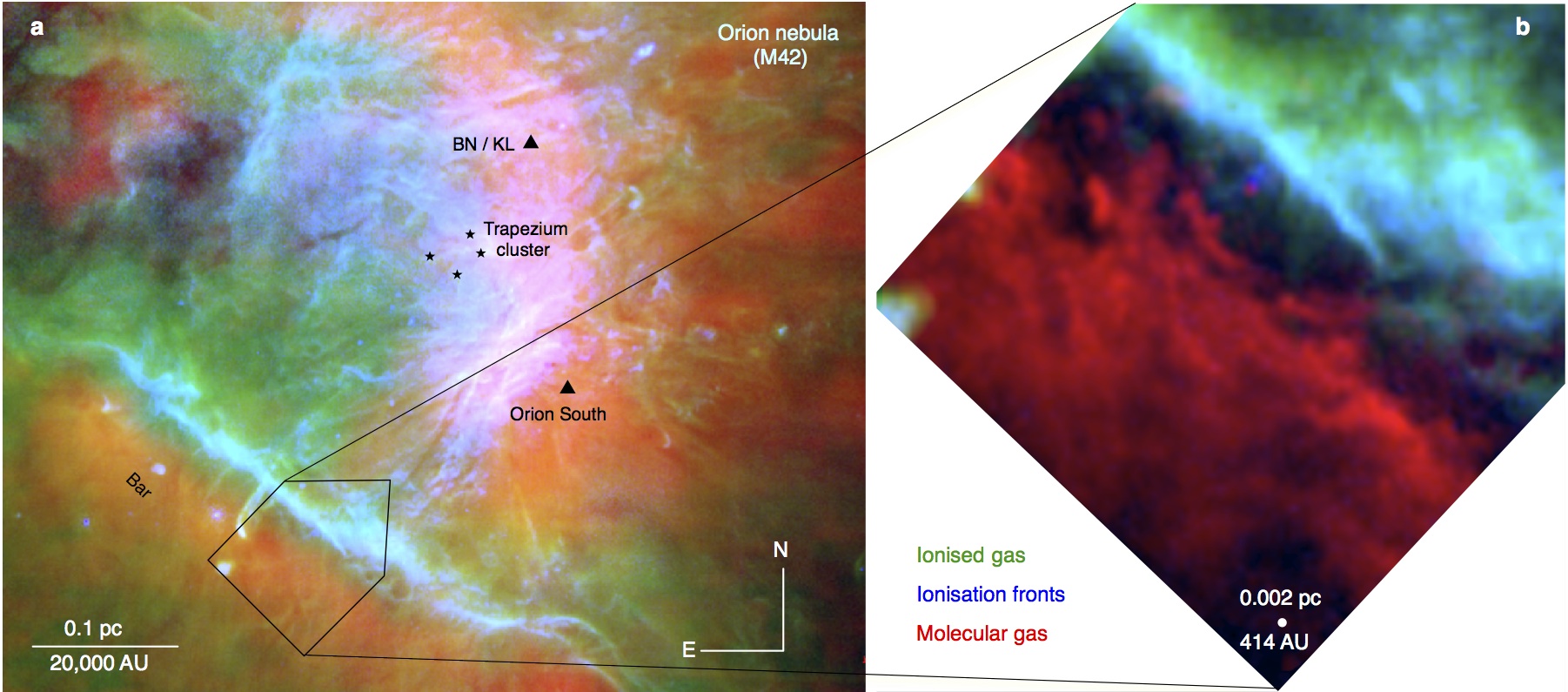}
\caption{\textbf{Multiphase view of the Orion nebula and molecular cloud. a},
Overlay of the HCO$^+$ $J$=3-2 emission (red) tracing the extended Orion molecular cloud. The hot ionised gas surrounding the Trapezium stars is shown by the [\SII]\,6,731\,\AA~emission (green). The interfaces between the ionised and the neutral gas, the ionisation fronts, are traced by the [\OI]\,6,300\,\AA~emission (blue), both lines imaged with VLT/MUSE\cite{Weilbacher15}. The size of the image is $\sim$5.8$'$$\times$4.6$'$. \textbf{b}, Close-up of the Bar region imaged with ALMA in the \mbox{HCO$^+$ $J$=4-3} emission (red). The black-shaded region is the atomic layer.}
\end{figure}

\begin{figure}
\includegraphics[width=1\textwidth]{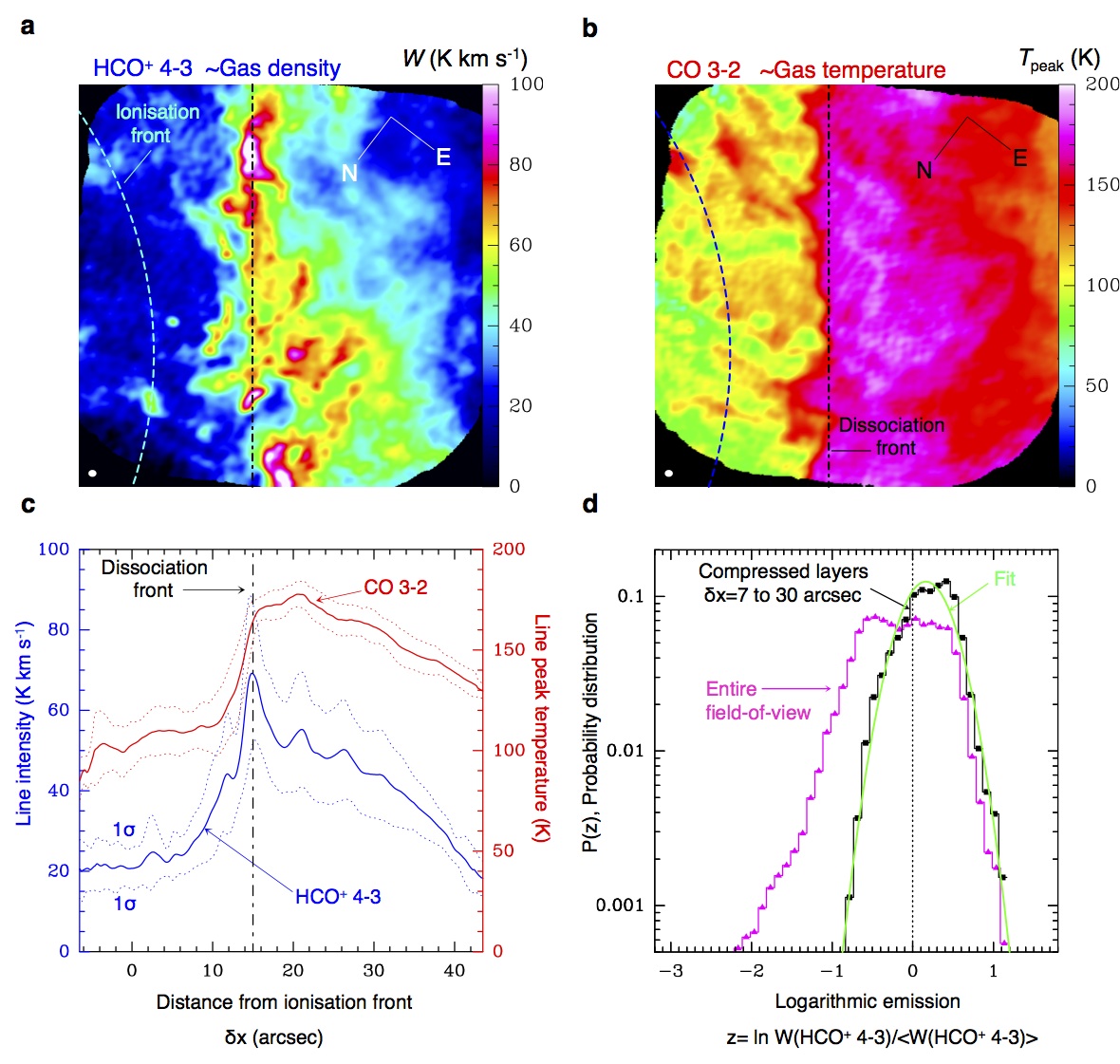}
\caption{\textbf{ALMA images of the Orion Bar. a}, 
HCO$^+$ $J$=4-3 line integrated intensity. \textbf{b}, CO $J$=3-2 line peak. Compared to Fig.~1, images \textbf{a} and \textbf{b} have been rotated by 127.5$^o$ anticlockwise to bring the incident UV radiation from left (see sketch in Extended Fig.~1). The dashed curve and the vertical dotted-dashed line delineate the ionisation and dissociation fronts respectively\cite{Walmsley00}. \textbf{c}, Vertically-averaged intensity cuts perpendicular to the Bar in $W_{4-3}^{{\rm HCO^+}}$ (blue curve) and $T_{\rm peak}^{{ \rm CO\,3-2}}$ (red curve). \textbf{d}, Probability distribution of $W_{4-3}^{{\rm HCO^+}}$ (proportional to the gas density) in the observed field (magenta triangles) and in the compressed layers (black squares).}
\end{figure}

\newpage
\section*{Methods}

\subsection*{ALMA interferometric and IRAM-30m single-dish observations}
ALMA Cycle-1 observations of the Bar were carried out using twenty-seven \mbox{12 m} antennae in band 7 at 345.796\,GHz (CO $J$=3-2) and 356.734\,GHz (HCO$^+$ $J$=4-3).  The observations consisted of a 27-pointing mosaic centred at \mbox{$\alpha$(2000) = 5$^h$35$^m$20.6$^s$};  \mbox{$\delta$(2000) = -05$^o$25$'$20$''$}. The total field-of-view (FoV) is 58$''$$\times$52$''$. Baseline configurations from $\sim$12 to $\sim$444\,m were used  (C32-3 antennae configuration). Lines were observed with correlators providing $\sim$500\,kHz resolution ($\delta$v$\approx$0.4~km\,s$^{-1}$) over a 937.5~MHz bandwidth. The ALMA 12 m array total observation time was $\sim$2h. ALMA executing blocks were first calibrated in the CASA software (version 4.2.0) and then exported to GILDAS.  In order to recover the large-scale extended emission filtered out by the interferometer, we used fully sampled single-dish maps as ``zero-'' and ``short-spacings''. Maps were obtained with the  IRAM-30m telescope (Pico Veleta, Spain) using the EMIR330 receiver under excellent winter conditions ($<$1~mm of precipitable water vapour). On-The-Fly (OTF) scans of a 170$''$$\times$170$''$ region were obtained along and perpendicular to the Orion Bar. The beam full-width at half maximum power  (FWHM) at 350GHz is $\sim$7$''$.  The GILDAS/MAPPING software was used to create the short-spacing visibilities[31] not sampled by ALMA. These visibilities were merged with the interferometric observations. Each mosaic field was imaged and a dirty mosaic was built. The dirty image was deconvolved using the standard H{\"o}gbom CLEAN algorithm and the resulting cubes were scaled from Jy/beam to brightness temperature scale using the synthesized beam size of $\sim$1$''$. This is a factor of $\sim$9 higher resolution than previous interferometric observations of the HCO$^+$ $J$=1-0 line towards the Bar\cite{Young00}.  
The achieved rms noise is $\sim$0.4~K per
 0.4 km~s$^{-1}$ channel, with an absolute flux accuracy of $\sim$10\%. The resulting images are shown in Fig.~1b and 2, and in Extended Data Fig.~2.   Finally, the large-scale HCO$^+$ $J$=3-2 (267.558 GHz) OTF map shown in Fig.~1a was taken with the multi-beam receiver HERA, also at the IRAM-30m telescope. The spectral and angular resolutions are $\sim$0.4~km\,s$^{-1}$ and 9$''$ (FWHM) respectively. The final images were generated using the GILDAS/GREG software.

\subsection*{Saturation and extinction corrections for the near-infrared image}

To better understand the spatial distribution of the H$_2$ $v$=1-0 $S$(1) line emission at $\lambda$=2.12~$\mu$m (H$_{2}^{*}$) presented in ref.\cite{Walmsley00} and shown in Extended Data Fig.~2, we note two effects that determine the resulting emission morphology. First, there is a bright star in the line of sight towards the Bar 
($\Theta^2$AOri at $\alpha$(2000)=5$^h$35$^m$22.9$^s$;  $\delta$(2000)=-05$^o$24$'$57.8$''$) that saturates the near-IR detectors in a slit of $\sim$4$''$-width parallel to the Bar (roughly between $\delta x =19''$ and 23$''$ in our rotated images). Hence, no H$_{2}^{*}$  data is shown in this range. Therefore, the layers with H$_2$ vibrational emission are wider that what Extended Data Fig.~2 might suggest, and more H$_{2}^{*}$ emission peaks can coincide with HCO$^+$ peaks in the blanked $\delta x =19''-23''$ region. Older, lower-angular and -spectral resolution near-IR images do show[32] that the H$_{2}^{*}$ emission extends out to $\delta x \simeq 20''$. Second, dust extinction (due to foreground dust in Orion's Veil and also due to dust in the Bar itself) may affect the apparent morphology of the near-IR images. Such effects are often neglected[1, 32, 33] and are not included in Extended Data Fig.~2. The extinction towards the Bar produced by the Veil is not greater than about 2 mag[34].  Adopting a dust reddening appropriate to Orion[11, 35], \mbox{$R_{\rm V}$=$A_{\rm V}$/$E$(B-V)=5.5}, and the 
\mbox{$A_{\rm K}$/$A_{\rm V}$} extinction law in ref.[35], we estimate that the H$_{2}^{*}$ emission lines would only be a factor of $\sim$30\%~brighter if foreground extinction corrections are taken into account.  An additional magnitude of extinction due to dust in the atomic layer of the Bar itself, results in a line intensity increase of $\sim$50\%. Therefore, minor morphological differences between the near-IR and millimetre-wave images can reflect a small-scale or patchy extinction differences in the region\cite{Walmsley00}. 

\subsection*{Excitation and radiative transfer models for CO and HCO$^+$} 

In order to estimate the physical conditions of the HCO$^+$ emitting gas near the dissociation front we run a grid of non-local, non-LTE excitation and radiative transfer (Monte Carlo) models.  This approach allows us to explore different column densities, gas temperatures and densities. Compared to most PDR models (using local escape probability approximations) our models take radiative pumping, line trapping and opacity broadening into account. This allows for the treatment of optically thick lines (see the Appendix in ref.[36] for code details and benchmarking tests). Our models use the most recent inelastic collisional rates of HCO$^+$ with H$_2$ and with electrons, and of CO with both H$_2$ and H. The electron density, $n_e$, can be an important factor in the collisional excitation of molecular cations in FUV-illuminated gas. For HCO$^+$, collisions with electrons start to contribute above $n_e$$>$10~cm$^{-3}$ 
(or $n_{\rm H}$$>$10$^5$~cm$^{-3}$ if most of the electrons are provided by carbon atom ionisation). In PDRs, collisions of molecules with H atoms can also contribute because the molecular gas fraction, $f$=2$n$(H$_2$)/$n_{\rm H}$=2$n$(H$_2$)/[$n$(H)+2$n$(H$_2$)], is not 1 (fully molecular gas). We adopted $f$=0.8 and varied $x_e$ between 0 and 10$^{-4}$. The H$_2$  ortho-to-para ratio was computed for each gas temperature $T$. Radiative excitation by the cosmic microwave background ($T_{\rm CMB}$=2.7~K) and by the FIR dust continuum in the Bar[37] (simulated by optically thin thermal emission at  $T_{\rm dust}=55~K$) were also included. 

Column densities of $N$(HCO$^+$)=(5$\pm$1)$\times$10$^{13}$~cm$^{-2}$ and $N$(CO)=(1.0$\pm$0.5)$\times$10$^{18}$~cm$^{-2}$ were estimated utilizing information from our IRAM-30m telescope line-survey towards the dissociation front[38]. This includes several HCO$^+$, H$^{13}$CO$^+$, HC$^{18}$O$^+$ and C$^{18}$O rotational lines in the estimation (the quoted dispersions in the column densities reflect the uncertainty obtained from least square fits to rotational population diagrams). They are consistent with previous observations in the region\cite{Hogerheijde95,Young00}. Radiative transfer models were run for $N$(HCO$^+$)=5$\times$10$^{13}$~cm$^{-2}$, $N$(CO)=1.0$\times$10$^{18}$~cm$^{-2}$, and $N_{\rm H}$=$N$(H)+2$N$(H$_2$)$\approx$2$\times$10$^{22}$~cm$^{-2}$ (equivalent to 
$A_{\rm V}$$\approx$7~mag for the dust properties in Orion). This results in $x$(HCO$^+$)$\approx$(2-3)$\times$10$^{-9}$ and $x$(CO)$\approx$(2.5-7.5)$\times$10$^{-5}$ abundances. In addition, the HCO$^+$/H$^{13}$CO$^+$ column density ratios derived from single-dish observations are similar to the $^{12}$C/$^{12}$C=67 isotopic ratio in Orion[39]. Thus, the H$^{12}$CO$^+$ lines are not very opaque ($\tau_{\rm line}$$\simeq$2) otherwise the observed HCO$^+$/H$^{13}$CO$^+$ line intensity ratios would be considerably smaller. A non-thermal (turbulent) velocity dispersion $\sigma_{\rm nth}$ of about 1~km\,s$^{-1}$ reproduces the observed line widths.  A similar value, 
$\sigma_{\rm nth}$$\simeq$1.0-1.5~km\,s$^{-1}$, is inferred directly from the observed line profiles ($\sigma_{\rm nth}^{2}$=$\sigma_{\rm obs}^2$--$\sigma_{\rm th}^{2}$, with
$\Delta$v$_{\rm FWHM}$=2$\sqrt{2ln2}$\,$\sigma_{\rm obs}$$\approx$3.0$\pm$0.5~km\,s$^{-1}$
and $T$=300~K). Hence, opacity broadening plays a minor role. The dispersion $\sigma_{\rm nth}$ is similar or lower than the local speed of sound at $T$=100-300~K 
($c_{\rm PDR}$=($k$$T$/$m$)$^{1/2}$=1.0-1.7~km\,s$^{-1}$, where $m$ is the mean mass per particle). This results in moderate Mach numbers 
($M$=$\sigma_{\rm nth}$/$c_{\rm PDR}\lesssim 1$).

Extended Data Fig.~3 shows model predictions for the CO $J$=3-2 line intensity peak, $T_{\rm peak}^{{ \rm CO\,3-2}}$ (upper left panel), and HCO$^+$ $J$=4-3 line integrated intensity, $W_{4-3}^{{\rm HCO^+}}$=$\int{T_B}$$d$v  (K\,km\,s$^{-1}$), for different $T$ and $n_{\rm H}$ values. For optically thick lines ($\tau_{\rm line}$$\gg$1),
$T_{\rm peak}^{{ \rm CO\,3-2}}$  provides a good measure of $T_{\rm ex}$, with \mbox{$T_{\rm peak}$$\approx$$J$($T_{\rm ex}$)=$E_{\rm up}$/$k$\,(e$^{E_{\rm up}/k\,T_{\rm ex}}-1$)$^{-1}$}. In addition, for low critical density ($n_{\rm cr}$) transitions such as the low-$J$ CO transitions, lines are close to thermalisation at densities above 
$\sim$10$^4$~cm$^{-3}$, thus $T_{\rm ex}$$\rightarrow$$T$ (with $n_{\rm cr}$=$A_{ij}$/$\gamma_{ij}$, where $A_{ij}$ is the Einstein coefficient for spontaneous emission and $\gamma_{ij}$ is the collisional de-excitation rate coefficient). 

In this case, $T_{\rm peak}^{{ \rm CO3-2}}$ is a good thermometer of the 
$\tau_{\rm CO 3-2}$$\gg$1 emitting layers. The HCO$^+$ $J$=4-3 line, however, has much higher critical densities ($n_{\rm cr,H_2}$$>$5$\times$10$^6$\,cm$^{-3}$ and $n_{\rm cr,e}$$\simeq$10$^3$~$e$\,cm$^{-3}$). For $n_{\rm H}$$<$2$n_{\rm cr,H_2}$/$\tau_{\rm line}$ (sub-thermal excitation), the integrated line intensity $W_{4-3}^{{\rm HCO^+}}$ is approximately linearly proportional to $N$(HCO$^+$)=$x$(HCO$^+$)$\cdot$$n_{\rm H}$$\cdot$$l$ even if the line is moderately thick.  
PDR models\cite{Young00,Sternberg95} and CO observations respectively show that $x$(HCO$^+$) and $T$ do not change significantly in the PDR layers around the H$_{2}^{*}$ emission peaks (between  $A_{\rm V}$$\approx$1 and 2~mag).  In a nearly edge-on PDR, the spatial length along the line of sight $l$ does not change much either. We compute that for the inferred $T$ and $N$(HCO$^+$) values in the region, the integrated line intensity 
$W_{4-3}^{{\rm HCO^+}}$ is proportional to density in the $n_{\rm H}$=10$^{4-6}$~cm$^{-3}$  range (the correlation coefficient is $r$$\simeq$0.98 for models with $x_e$=0 and 10$^{-4}$). Moreover, $W_{4-3}^{{\rm HCO^+}}$ still increases with density up to several 10$^6$~cm$^{-3}$ ($r$$\simeq$0.94). This reasoning justifies the use of $W_{4-3}^{{\rm HCO^+}}$ as a proxy for $n_{\rm H}$ in the region. 

\subsection*{Average physical conditions in the compressed structures}

The physical conditions that reproduce the mean CO~$J$=3-2 line peak and HCO$^+$~$J$=4-3 integrated line intensity towards the compressed structures at $\delta x \simeq 15''$ 
($T_{\rm peak}^{{ \rm CO\,3-2}}$=164$\pm$10~K and 
$W_{4-3}^{{\rm HCO^+}}$=69$\pm$18~K\,km\,s$^{-1}$) are $T$=200-300~K and $n_{\rm H}$=(1.0$\pm$0.5)$\times$10$^6$~cm$^{-3}$ (Extended Data Fig.~3).  This implies high thermal pressures, $P_{\rm th,comp}$/$k$=$n_{\rm H}$$\cdot$$T$$\approx$(1.0--4.5)$\times$10$^8$~K\,cm$^{-3}$. 
The brightest HCO$^+$ emission peaks (with $W_{4-3}^{{\rm HCO^+}}$$\simeq$100~K\,km\,s$^{-1}$, Fig. 2a) likely correspond to specific density enhancements. For the range of column densities and physical conditions at $\delta x \simeq 15''$, the $T$ uncertainty is determined by the lack of higher-$J$ CO lines, observed at high-angular resolution, to better constrain $T$ from models. The range of estimated gas densities is dominated by the dispersion ($\sim$25\%) of the mean $W_{4-3}^{{\rm HCO^+}}$. 

The above conditions suggest that the cloud edge contains substructures that are denser than the atomic layer ($n_{\rm H}$=(4-5)$\times$10$^4$~cm$^{-3}$)\cite{Andree-Labsch14,Tielens93} and denser than the ambient molecular cloud
($n_{\rm H}$=(0.5-1.0)$\times$10$^5$~cm$^{-3}$)\cite{Hogerheijde95}. The equivalent length of the substructures is small, $l$=$N_{\rm H}$/$n_{\rm H}$$\approx$(4-12)$\times$10$^{-3}$~pc ($\approx$2$''$-6$''$ at the distance to Orion, thus consistent with their apparent size in the ALMA image). The mass of a cylinder with $n_{\rm H}$ of a few 10$^6$~cm$^{-3}$, 2$''$-6$''$ length and width of 2$''$ is  $\lesssim$0.005~$M_{\odot}$  (a mass per unit length of  0.3-1.0~$M_{\odot}$~pc$^{-1}$). This is much lower than the Virial and critical  masses[40] needed to make them gravitationally unstable (about 5~$M_{\odot}$, from the inferred gas $T$, $n_{\rm H}$ and velocity dispersion).  H$_2$ clumps of similar small masses (several 0.001\,$M_{\odot}$) have been intuited towards the boundary of more evolved and distant \HII~regions[41].  Compression and fragmentation of UV-irradiated cloud edges must be a common phenomenon in the vicinity of young massive stars.

\subsection*{Physical conditions in the ambient molecular cloud}

Deeper inside the molecular cloud, $T_{\rm peak}^{{ \rm CO\,3-2}}$ smoothly decreases from $\sim$170~K to $\sim$130~K. Therefore, these observations do not suggest temperature spikes at scales of a few arcseconds. Deeper inside the molecular cloud ($\delta x>30''$ in our rotated images), both $N$(H$_2$) and $N$(HCO$^+$) are expected to gradually increase[5, 7, 6, 37]. For the expected $N$(HCO$^+$)$\approx$2$\times$10$^{14}$~cm$^{-2}$ column density\cite{Hogerheijde95,Young00}, excitation models show that the gas density in the ambient cloud is  $n_{\rm H}$$\approx$(0.5-1.0)$\times$10$^5$\,cm$^{-3}$ (dashed curves in  Extended Data Fig.~3), in agreement with previous estimations\cite{Tielens85,Hogerheijde95}. Hence, the over-dense substructures have compression factors $\sim$5-30 with respect to the ambient molecular gas.

\subsection*{Physical conditions in the atomic layer}

The decrease of both $T_{\rm peak}^{{ \rm CO\,3-2}}$  and $W_{4-3}^{{\rm HCO^+}}$ between the ionisation and dissociation fronts is consistent with the expected sharp decrease of CO and HCO$^+$ abundances in the atomic layer. The representative gas density in the atomic layer, $n_{\rm H}$$\approx$(4-5)$\times$10$^4$~cm$^{-3}$, is constrained by the strength of the un-attenuated FUV flux at the Bar edge\cite{Hogerheijde95,Andree-Labsch14} ($\chi$$\approx$4.4$\times$10$^{4}$, determined by the spectral type of the Trapezium stars) and by the current position of the dissociation front at $\delta x \simeq 15''$ [1, 33]. The exact gas density value, however, depends on the assumed FUV-extinction grain properties (which likely vary as function of cloud depth). In the context of stationary PDR models, larger-than-standard-size grains (lower FUV absorption cross-sections) are often invoked[33], otherwise the separation between the dissociation and ionisation fronts would be  smaller than the observed $\sim$15$''$. The lower densities in the atomic layer agree with the observed low H$_2$ $v$=1-0 $S$(1)/$v$=2-1 $S$(1)$\approx$3 line intensity ratio attributed to fluorescent H$_{2}^{*}$ excitation[32, 42]. We note that optically thin CO emission implies $T_{\rm peak}^{{ \rm CO\,3-2}}$$\ll$$T_{\rm ex}$. Hence,
$T_{\rm peak}^{{ \rm CO\,3-2}}$ can no longer be used as a gas thermometer in the atomic layer where the CO abundance is low. The gas temperature close to the dissociation front is between $T$$\approx$500~K (from \HI~observations\cite{Werf13}) and $T$$\approx$300~K (from carbon radio-recombination[43] and [\CII]158$\mu$m\,\cite{Goicoechea15} line observations). 

\subsection*{Emission Probability Distribution Functions (PDF)}

In order to study the distribution of gas densities in the region, approximated by the HCO$^+$ $J$=4-3 emission, we analysed the probability distribution of the logarithmic emission, given by  $z$=ln⁡($W_{4-3}^{{\rm HCO^+}}$/$<$$W_{4-3}^{{\rm HCO^+}}$$>$), where                     $<$$W_{4-3}^{{\rm HCO^+}}$$>$=$<$$\int{T_B}$\,$d$v$>$  is the mean value in the observed FoV (37~K\,km\,s$^{-1}$). This is a common approach to interpret (column) density maps, both from observations and MHD simulations[24, 44]. The PDF is computed as the number of pixels (in the high signal-to-noise $W_{4-3}^{{\rm HCO^+}}$ image) per intensity bin divided by the total number of pixels. We first analysed the complete FoV observed by ALMA and selected $W_{4-3}^{{\rm HCO^+}}$ measurements above 5 sigma, where we define \mbox{sigma=rms$\cdot$(2$\delta$v$\cdot$$\Delta$v$_{\rm FWHM})^{1/2}$}, with 
$\delta$v=0.4~km\,s$^{-1}$ and  v$_{\rm FWHM}$=3.0~km\,s$^{-1}$. The resulting PDF is shown in Fig.~2d  (magenta points). Second, we selected measurements only in the compressed layers region between $\delta x$=7$''$ and 30$''$ (with respect to the rotated images in Fig.~2). The resulting PDF (black points) is very close to a log-normal distribution with $p(z)=N exp(-(z-z_0)^2/2\sigma^2)$, where $z_0$ is the peak value 
and $\sigma$ the standard deviation. We obtain $z_0$=0.165 and $\sigma$=0.31 from a fit (green curve). If  $W_{4-3}^{{\rm HCO^+}}$ is proportional to the gas density, these values imply that 99\%~of the observed positions in the compressed layers span a factor of $\sim$6 in density. In MHD models, $\sigma$ is a measure of how density varies in a turbulent cloud. Hence, it depends on the Mach number, the ratio of the thermal to magnetic pressure ($\beta$) and the forcing characteristics of the turbulence\cite{Federrath13}. The relatively modest $\sigma$ value inferred in the $\delta x$=7$''$-30$''$ layer is consistent with the low Mach numbers in the PDR, and suggests a significant role of magnetic pressure. We note that a similar analysis of the CO emission does not yield the same log-normal distribution. This is consistent with low-$J$ CO lines being optically thick and tracing gas temperature and not gas density variations. This reinforces that the log-normal shape of the $W_{4-3}^{{\rm HCO^+}}$ PDF in the compressed layer is a relevant observational result. 

\subsection*{Gas pressures, magnetic field and compression}

To support to the cloud compression and gas photo-ablation scenario, we investigated the different contributions to the gas pressure in the region. The thermal pressure in the \HII~region near the ionisation front\cite{Walmsley00} is $P_{\rm th,HII}$/$k$=2$\cdot$$n_e$$\cdot$$T_e$$\approx$6$\times$10$^7$~K\,cm$^{-3}$, about 6 times higher than the turbulent ram pressure $P_{\rm ram,amb}$=$\rho$\,$\sigma_{\rm nth,amb}^{2}$ in the ambient molecular cloud (Extended Data Table~1). Since we find similar contribution of thermal and non-thermal (turbulent) pressures in both the ambient cloud and the over-dense substructures 
($\alpha$=$P_{\rm nth,amb}$/$P_{\rm th,amb}$$\approx$$P_{\rm nth,comp}$/$P_{\rm th,comp}$$\approx$1), it is reasonable to assume equipartition of thermal, turbulent and magnetic energies to quantify the magnetic pressure in the PDR ($P_B$=$B^2$/8$\pi$). In particular, for $\beta$=$P_B$/$P_{\rm th}$=1 we estimate magnetic fields strengths of $B$$\approx$200~$\mu$G and $\approx$800\,$\mu$G in the ambient and in the high-density substructures respectively. Such strong magnetic fields at small scales need to be confirmed observationally (both the strength and the orientation) but seem consistent with the high values ($\sim$100\,$\mu$G) measured in the low-density foreground material[45] (the ``Orion Veil'') confirming that $B$ is particularly strong in the Orion complex. On much larger spatial scales, low-angular resolution observations do suggest that B increases with gas density at \HII/cloud boundaries ($B \propto n_{\rm H}^{0.5-1}$)[46].

      A strong magnetic field would be associated with large magnetosonic speeds 
(v$_{\rm ms}$) in the PDR.  If a UV-driven shock-wave is responsible for the molecular gas compression, its velocity is predicted to slow down to v$_{\rm s}$$\lesssim$3~km\,s$^{-1}$ once entered the molecular cloud\cite{Hill78}. In such a slow, magnetised shock 
(v$_{\rm s}$$\ll$v$_{\rm ms}$), compression waves can travel ahead of the shock front[47]. Thus, a high magnetic field strength may be related with the $W_{4-3}^{{\rm HCO^+}}$ undulations seen perpendicular to the Bar (Fig.~2c). The inferred compression factor in the observed substructures ($f$=$n_{\rm comp}$/$n_{\rm amb}$=5-30) is consistent with slow shock velocities\cite{Draine11}, 
 v$_{\rm s}$=$c_0$$\sqrt{f}$$\approx$1.5-4.0~km\,s$^{-1}$, where $c_0$ is the initial sound speed of the unperturbed molecular gas. The necessarily small v$_{\rm s}$ agrees with the relatively narrow molecular line-profiles ($\Delta$v$_{\rm FWHM}$$\lesssim$4~km\,s$^{-1}$) seen in PDRs\cite{Hollenbach99} (including observations of face-on sources in which the shock would propagate in the line of sight). Owing to the high thermal pressure in the compressed structures, we also find that a pressure gradient, with $P_{\rm th,comp}$$\geq$$P_{\rm th,HII}$ exists. This subtle effect is seen in simulations of an advancing shock-wave around an \HII~region[22, 48].  

\subsection*{Molecular gas between the ionization and dissociation fronts}

ALMA reveals fainter HCO$^+$ and CO emission in the atomic layer (HCO$^+$ globulettes and plume-like CO features at $\delta x <15''$, Fig.~2). Previous low-angular resolution observations and models had suggested the presence of dense spherical clumps with 5$''$-10$''$ sizes deeper inside the molecular cloud\cite{Young00,Lis03} (at $\geq$15$''$-20$''$ from the ionisation front[3, 6, 32]). The dense substructures resolved by ALMA are smaller ($\simeq$2$''$$\times$4$''$) and are detected at $\delta x \gtrsim 7''$ (even before the peak of the H$_2$ vibrational emission).

The molecular line profiles towards the plumes typically show two velocity emission components (Extended Data Fig.~4). One centred at v$_{\rm LSR}$$\approx$8.5~km\,s$^{-1}$, the velocity of the background molecular cloud in the back-side of M42\cite{Goicoechea15} (not directly associated with the Bar), and other at  v$_{\rm LSR}$$\approx$11~km\,s$^{-1}$, the velocity-component of the molecular gas in the Bar. In addition, despite the small size of the observed region, the crosscuts of the HCO$^+$ $J$=4-3 line velocity centroid and of the FWHM velocity dispersion, show gradients perpendicular to the Bar (Extended Data 
Fig.~4). Moving from the ionisation front to the molecular gas, the line centroid shifts to higher velocities (gas compression effects may, in part, contribute to this red-shifted velocity). The velocity dispersion, however, shows its maximum between the ionisation and the dissociation fronts, the expected layers for photoablative neutral gas flows. Both the kinematic association with the Bar velocities and the higher velocity dispersion between the two fronts is are consistent with the presence of gas flowing from the high-pressure compressed molecular layers 
($P_{\rm th,comp}$/$k$$\approx$2$\times$10$^8$~K\,cm$^{-3}$) to the atomic layers ($P_{\rm th,atomic}$/$k$$\approx$5$\times$10$^7$~K\,cm$^{-3}$).

\subsection*{HCO$^+$ chemistry and the C$^+$/CO transition zone}

Static equilibrium PDR models appropriate to the ambient gas component 
($n_{\rm H}$$\approx$5$\times$10$^4$\,cm$^{-3}$) reproduce the separation between the ionisation and dissociation fronts\cite{Young00}. However, they predict HCO$^+$ abundances near the dissociation front that are too low ($x$(HCO$^+$) of a few 5$\times$10$^{-11}$) to be consistent with the bright ridge of HCO$^+$ emission detected by ALMA. These models also predict that the C$^+$/CO transition should occur ahead of the H/H$_2$ transition zone and deeper inside the molecular cloud (at $\delta x \simeq 20''$’ from the ionisation 
front\cite{Andree-Labsch14,Tielens93}). However, our detection of bright CO and HCO$^+$ emission towards the layers of bright H$_2$ vibrational emission\cite{Walmsley00} implies that the C$^+$/CO transition occurs closer to the cloud edge, and nearly coincides with the H/H$_2$ transition (at least it cannot be resolved at the $\sim$1$''$ resolution of our observations).  This is likely another signature of dynamical effects. Indeed, the presence of molecular gas near the cloud edge[49], and a reduced C$^+$ abundance deeper inside the molecular cloud[50], may explain model and observation discrepancies of other chemically related molecules. 
       As an example, stationary PDR models applied to the fluorine chemistry[51] overpredict the CF$^+$ column density observed towards the Bar[52] by a factor of $\sim$10. Given that HF readily forms as F atoms react with H$_2$ molecules, CF$^+$ must arise from layers where C$^+$ and H$_2$ overlap (CF$^+$ forms through 
\mbox{HF + C$^+$ $\rightarrow$ CF$^+$ + H} reactions and is quickly destroyed by recombination with electrons)[51, 53]. Hence, the (lower-than-predicted) observed
 CF$^+$ abundances likely reflect a dynamical PDR behaviour as well. 

Stationary PDR models of strongly irradiated dense gas (with $n_{\rm H}$ of a few 
10$^6$~cm$^{-3}$) have been presented in the literature\cite{Andree-Labsch14,Young00,Sternberg95}. The above densities are similar to those inferred in the compressed substructures at the Bar edge. Thus they can be used to get insights about the chemistry that leads to the formation of HCO$^+$ and CO in UV-irradiated dense gas. Owing to the higher densities and enhanced H$_2$ collisional de-excitation heating, the gas attains high temperatures. This triggers a warm chemistry in which endothermic reactions and reactions with energy barriers become faster. As a result, higher HCO$^+$ abundances are predicted close to the dissociation front ($x$(HCO$^+$) of several 10$^{-9}$). Reactions of C$^+$ with H$_2$ (either far-UV-pumped or thermally excited) initiate the carbon chemistry[54]. This reaction triggers the formation of CH$^+$ (explaining the elevated CH$^+$ abundances detected by Herschel[55]) and reduces the abundance of C$^+$ ions and H$_2$ molecules near the dissociation front; i.e., the H/H$_2$ and the C$^+$/CO transition layers naturally get closer 
(in $A_{\rm V}$)[50]. Fast exothermic reactions of CH$^+$ with H$_2$ subsequently produce CH$_{2}^{+}$ and CH$_{3}^{+}$. Both hydrocarbon ions are ``burnt'' in reactions with abundant oxygen atoms and contribute to the HCO$^+$ formation at the molecular cloud edge. This HCO$^+$ formation route from CH$^+$ can dominate over the formation of HCO$^+$ from CO$^+$ (after the \mbox{O + H$_2$ $\rightarrow$ OH + H} reaction, followed by 
\mbox{C$^+$ + OH $\rightarrow$ CO$^+$ + H}, and finally 
\mbox{CO$^+$ + H$_2$ $\rightarrow$ HCO$^+$ + H)} [5, 6, 32]. 
Both OH and CO$^+$ have been detected in the Bar[56, 57], but high-angular 
resolution maps do not exist. Recombination of HCO$^+$ with electrons then drives CO production near the dissociation front[6, 7].

 Extrapolating the above chemical scenario, the brightest HCO$^+$ $J$=4-3 emission peaks in the Bar should be close to H$_{2}^{*}$ emission peaks.  
 Extended Data Fig.~2a shows a remarkable spatial agreement between the H$_2$ $v$=1-0 $S$(1) emission peaks and several HCO$^+$ emission peaks. Detailed H$_2$ excitation models (including both far-UV-pumping and collisions) show that for the conditions prevailing in the Bar, the intensity of the H$_2$ $v$=1-0 $S$(1) line is approximately proportional to the gas density[42]. Therefore, the HCO$^+$ peaks that match the position of the H$_2$ $v$=1-0 $S$(1) line peaks likely correspond to gas density enhancements as well. This agrees with the higher H$_2$ $v$=1-0 $S$(1)/$v$=2-1 $S$(1)$\approx$8 line intensity ratios observed at the dissociation front and consistent with significant H$_2$ collisional excitation[32]. The ALMA images thus confirm that in addition, or as a consequence of dynamical effects, reactions of H$_2$ with abundant atoms and ions contribute to shift the molecular gas production towards the cloud edge. Even higher-angular resolution observations of additional tracers will be needed to fully understand this, and to spatially resolve the chemical stratification expected in the overdense substructures themselves. We note that if most carbon becomes CO at $A_{\rm V}$$\approx$2 
($N_{\rm H}$ of a several 10$^{21}$~cm$^{-2}$) in substructures with gas densities of a few 10$^6$~cm$^{-3}$, this depth is equivalent to a spatial length of several 
10$^{15}$~cm, or an angular-scale of $\sim$0.5$''$ at the distance to Orion.

 Deeper inside into the molecular cloud ($\delta x >30''$), the CO$^+$, CH$^+$, CH$_{2}^{+}$ and CH$_{3}^{+}$ abundances sharply decrease. The far-UV flux significantly diminishes, and the gas and dust grain temperatures accordingly decrease. The HCO$^+$ abundance also decreases until the  \mbox{CO + H$_{3}^{+}$ $\rightarrow$ HCO$^+$ + H$_2$} reaction starts to drive the HCO$^+$ formation at low temperatures. Gas-phase atoms and molecules gradually deplete and dust grains become coated by ices as the FUV photon flux is attenuated at even larger cloud depths (see sketch in Extended Data Fig. 1).

\clearpage

\begin{enumerate}
  \setcounter{enumi}{30}

\item	Pety, J. \& Rodr\'{\i}guez-Fernandez, N. J. Revisiting the theory of interferometric wide-field synthesis.  \emph{{Astron.  Astrophys.}} \textbf{517}, A12  (2010). 

\item	van der Werf, P. P., Stutzki, J., Sternberg, A.  \& Krabbe, A. Structure and chemistry of the Orion bar photon-dominated region. \emph{{Astron. Astrophys.}} \textbf{313}, 633-648 (1996).

\item	Allers, K. N., Jaffe, D. T., Lacy, J. H., Draine, B. T.  \& Richter, M. J. H2 Pure Rotational Lines in the Orion Bar. \emph{{Astrophys. J.}}  \textbf{630}, 368-380 (2005). 

\item	O'Dell, C. R. \& Yusef-Zadeh, F., High Angular Resolution Determination of Extinction in the Orion Nebula. \emph{{Astron. J.}}, \textbf{120}, 382-392 (2000).

\item	Cardelli, J. A. Clayton, G. C. \& Mathis, J. S. The relationship between infrared, optical, and ultraviolet extinction. \emph{{Astrophys. J.}}  \textbf{345}, 245-256 (1989).

\item	Goicoechea, J. R. \emph{et~al.} Low sulfur depletion in the Horsehead PDR. 
\emph{{Astron.  Astrophys.}}  456, 565-580 (2006). 

\item	Arab, H. \emph{et~al.} Evolution of dust in the Orion Bar with Herschel. I. Radiative transfer modelling. \emph{{Astron. Astrophys.}}  \textbf{541}, A19 (2012).

\item	Cuadrado, S. \emph{et~al.} The chemistry and spatial distribution of small hydrocarbons in UV-irradiated molecular clouds: the Orion Bar PDR.  \emph{{Astron. Astrophys.}} \textbf{575}, A82 (2015). 

\item	Langer, W. D.  \& Penzias, A. A. $^{12}$C/$^{13}$C isotope ratio across the Galaxy from observations of $^{13}$C$^{18}$O in molecular clouds. \emph{{Astrophys. J.}}  \textbf{357}, 477-492 (1990).

\item	Inutsuka, S.-i. \& Miyama, S. M.  A Production Mechanism for Clusters of Dense Cores. \emph{{Astrophys. J.}} \textbf{480}, 681-693 (1997).

\item	Noel, B. Joblin, C., Maillard, J. P.  \& Paumard, T.  New results on the massive star-forming region S106 by BEAR spectro-imagery. \emph{{Astron. Astrophys.}}  \textbf{436}, 569-584 (2005).

\item	Burton, M. G., Hollenbach, D. J. \& Tielens, A. G. G. M. Line emission from clumpy photodissociation regions. \emph{{Astrophys. J.}}  \textbf{365}, 620-639 (1990).

\item	Wyrowski, F., Schilke, P., Hofner, P.,  \& Walmsley, C. M. Carbon Radio Recombination Lines in the Orion Bar. \emph{{Astrophys. J.}}  Letters. \textbf{487}, L171-L174 (1997).

\item	Tremblin, P. \emph{et~al.}  Ionization compression impact on dense gas distribution and star formation. Probability density functions around \HII~regions as seen by Herschel. \emph{{Astron. Astrophys.}} \textbf{564}, A106 (2014).

\item	Brogan, C. L., Troland, T. H., Abel, N. P., Goss, W. M., \& Crutcher, R. M.  
\HI~and OH Zeeman Observations Toward Orion's Veil. \emph{{Astronomical Polarimetry: Current Status and Future Directions}} \textbf{343}, 183 (2005).

\item	Planck Collaboration, \emph{et~al.} Planck intermediate results. XXXIV. The magnetic field structure in the Rosette Nebula. \emph{{Astron. Astrophys.}} \textbf{586}, A137 (2016).

\item	Roberge, W. G., \& Draine, B. T.  A new class of solutions for interstellar magnetohydrodynamic shock waves. \emph{{Astrophys. J.}} \textbf{350}, 700-721(1990).

\item	Raga, A. C., Cant\'o, J., \& Rodriguez, L. F. Analytic and numerical models for the expansion of a compact \HII~region. \emph{{MNRAS.}} \textbf{419}, L39-L43 (2012)

\item	Hollenbach, D. J. \& Natta, A.  Time-Dependent Photodissociation Regions. \emph{{Astrophys. J.}}  \textbf{455}, 133-144  (1995).

\item	Bertoldi, F.  ISO: A Novel Look at the Photodissociated Surfaces of Molecular Clouds. \emph{{ESA Special Publication}} \textbf{419}, 67-72 (1997).

\item	Neufeld, D. A. \& Wolfire, M. G. The Chemistry of Interstellar Molecules Containing the Halogen Elements. \emph{{Astrophys. J.}} \textbf{706}, 1594-1604 (2009).

\item	Neufeld, D. A. \emph{et~al.}  Discovery of interstellar CF$^+$. \emph{{Astron. Astrophys.}} \textbf{454}, L37-L40 (2006).

\item	Guzm\'an, V. \emph{et~al.} The IRAM-30m line survey of the Horsehead PDR. I. CF$^+$ as a tracer of C$^+$ and as a measure of the fluorine abundance. 
\emph{{Astron. Astrophys.}} \textbf{443}, L1 (2012).

\item	Ag\'undez, M., Goicoechea, J. R., Cernicharo, J., Faure, A. \& Roueff, E. The Chemistry of Vibrationally Excited H$_2$ in the Interstellar Medium. 
\emph{{Astrophys. J.}} \textbf{713}, 662-670 (2010).

\item	Nagy, Z. \emph{et~al.}   The chemistry of ions in the Orion Bar I. - CH$^+$, SH$^+$, and CF$^+$. The effect of high electron density and vibrationally excited H$_2$ in a warm PDR surface. \emph{{Astron. Astrophys.}}  \textbf{550}, A96  (2013).

\item	Goicoechea, J. R. \emph{et~al.} OH emission from warm and dense gas in the Orion Bar PDR. \emph{{Astron. Astrophys.}}   \textbf{530}, L16 (2011).

\item	Stoerzer, H., Stutzki, J. \& Sternberg, A. CO$^+$ in the Orion Bar, M17 and S140 star-forming regions. \emph{{Astron. Astrophys.}} \textbf{296}, L9-L12 (1995).

\end{enumerate}

\clearpage
\section*{Extended Data}

\renewcommand{\figurename}{Extended Data Figure}
\setcounter{figure}{0}   

\begin{figure}[h]
\includegraphics[width=1\textwidth]{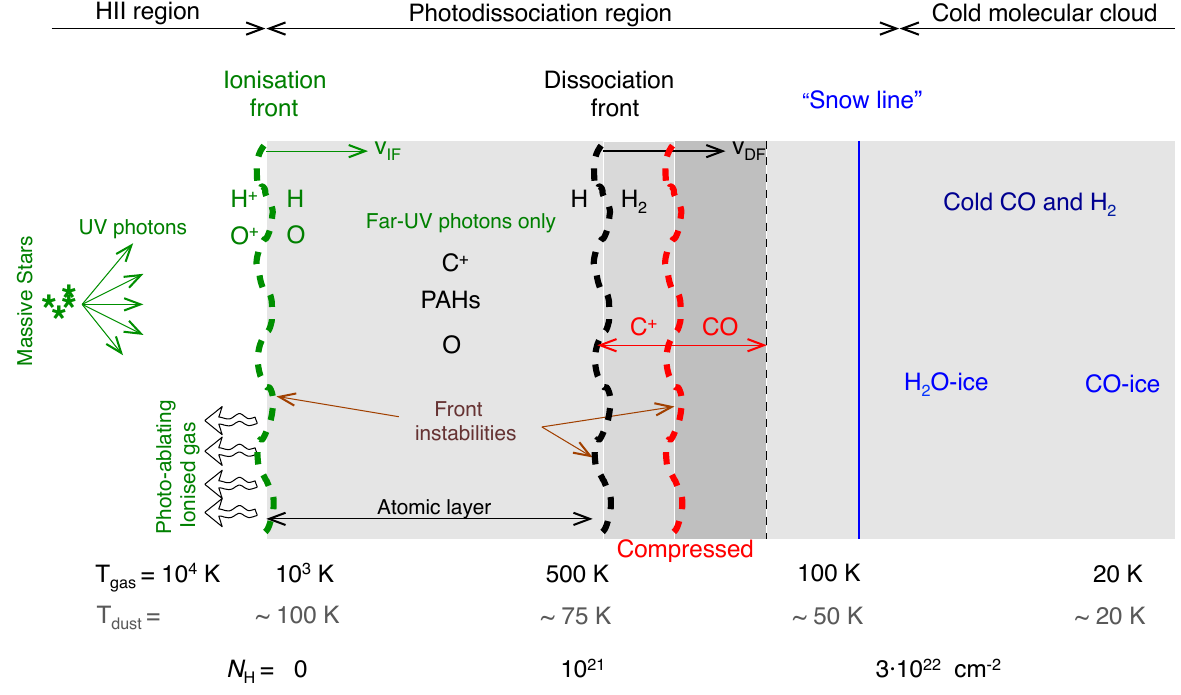}
\caption{\textbf{Structure of a strongly UV-irradiated molecular cloud edge}. 
The incident stellar UV radiation comes from the left. The velocity of the advancing ionisation and dissociation fronts are represented by v$_{\rm IF}$ and v$_{\rm DF}$ respectively. In the Orion Bar, the dissociation front is at about 15$''$ ($\sim$0.03~pc) from the ionisation front.}
\end{figure}

\begin{figure}[h]
\includegraphics[width=1\textwidth]{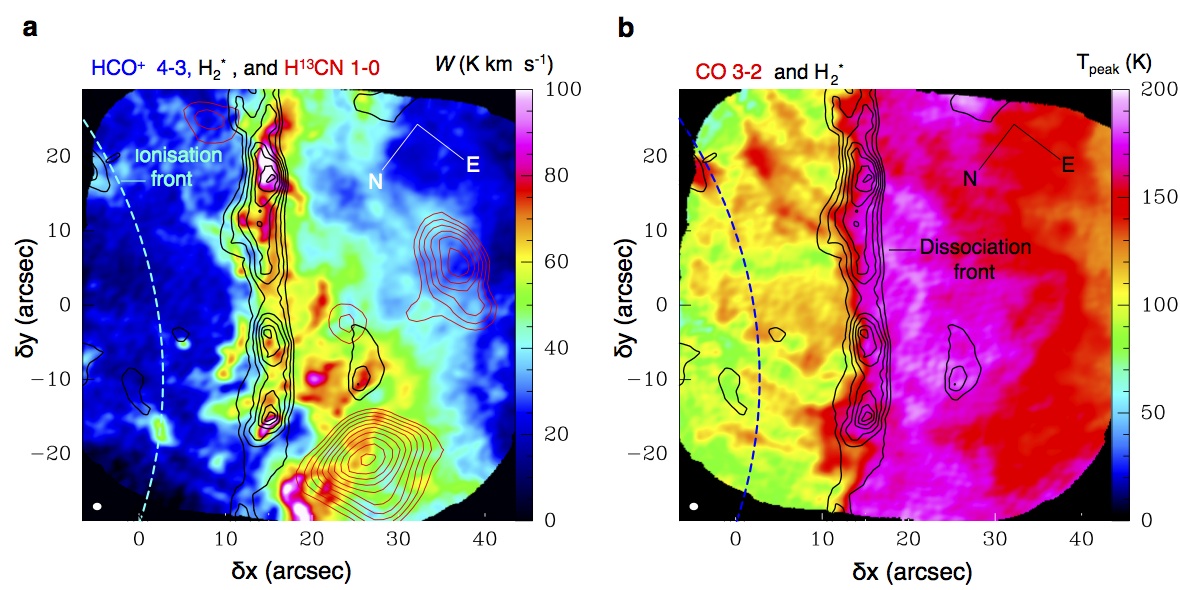}
\caption{\textbf{Comparison with other tracers.  a}, 
ALMA HCO$^+$ $J$=4-3 line integrated intensity,  \textbf{b}, ALMA CO $J$=3-2 line peak (Bar velocity component). The red contours represent the H$^{13}$CN $J$=1-0 emission (from 0.08 to 0.026 by 0.02~Jy~beam$^{-1}$~km~s$^{-1}$) of dense condensations inside the Bar	\cite{Lis03}. The black contours show the brightest regions of H$_2$ $v$=1-0 $S$(1) 
emission\cite{Walmsley00} (from 1.5 to 4.5 by 0.5$\times$10$^{-4}$~erg~s$^{-1}$~cm$^{-2}$~sr$^{-1}$). The H$_{2}^{*}$ image is saturated between $\delta$$x$=19$''$ and 23$''$ (i.e. no data is shown). Figures have been rotated by 127.5$^o$ anticlockwise to bring the incident UV radiation from left.}
\end{figure}

\begin{figure}[h]
\includegraphics[width=1\textwidth]{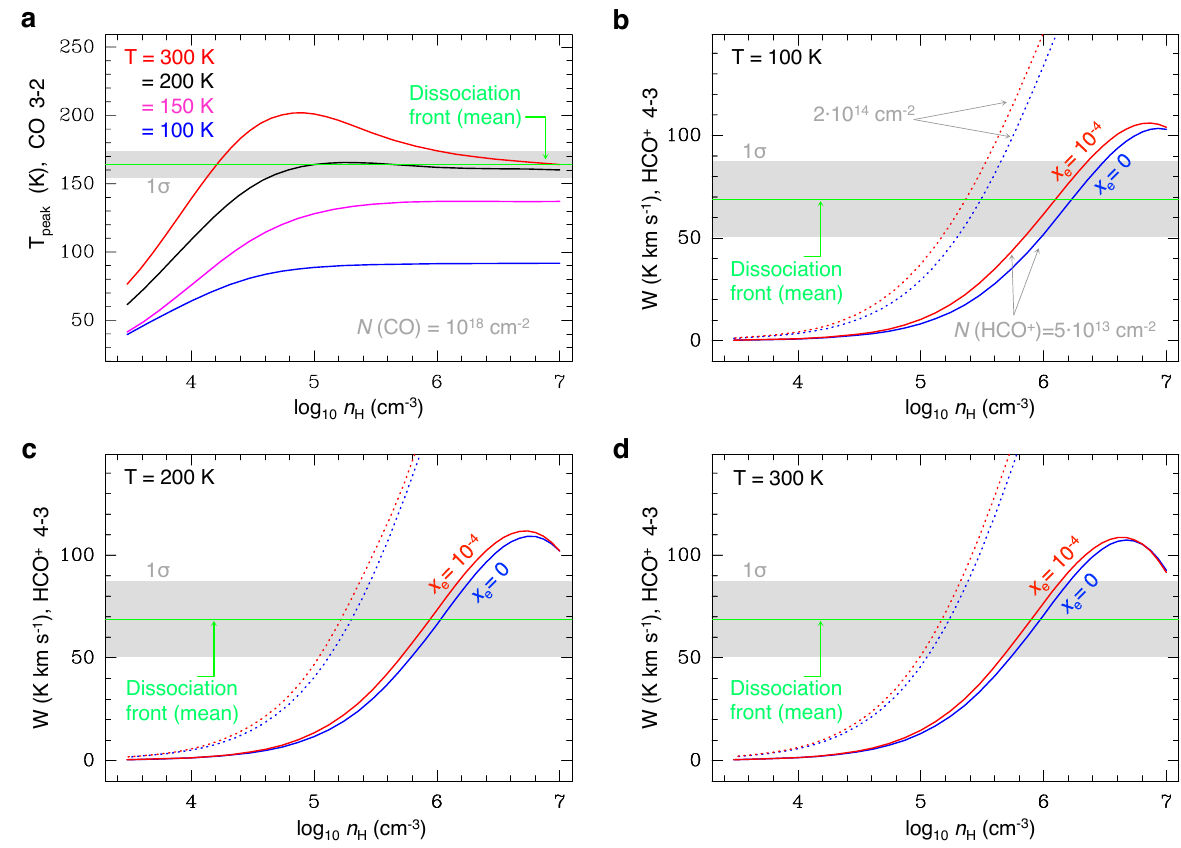}
\caption{\textbf{Excitation models for different gas temperatures and densities. a},
CO $J$=3-2 line peak (for $N$(CO)=10$^{18}$~cm$^{-2}$). \textbf{b}, \textbf{c} and \textbf{d} HCO$^+$ \mbox{$J$=4-3} integrated line intensity. Each curve represents a different electron abundance model: $x_e$=0 (blue) and 10$^{-4}$ (red).  Continuous curves are for $N$(HCO$^+$)=5$\times$10$^{13}$~cm$^{-2}$ and dotted lines for $N$(HCO$^+$)=2$\times$10$^{14}$~cm$^{-2}$ (appropriate for deeper inside the Bar, $\delta$$x$$>$30$''$).   The horizontal green dashed line represents the average 
$T_{\rm peak}^{{ \rm CO\,3-2}}$ (a) and $W_{4-3}^{{\rm HCO^+}}$ (b, c, and d) with their standard deviation (grey shaded) towards the dissociation front (at $\delta$$x$$\approx$15$''$).}
\end{figure}

\begin{figure}[h]
\includegraphics[width=1\textwidth]{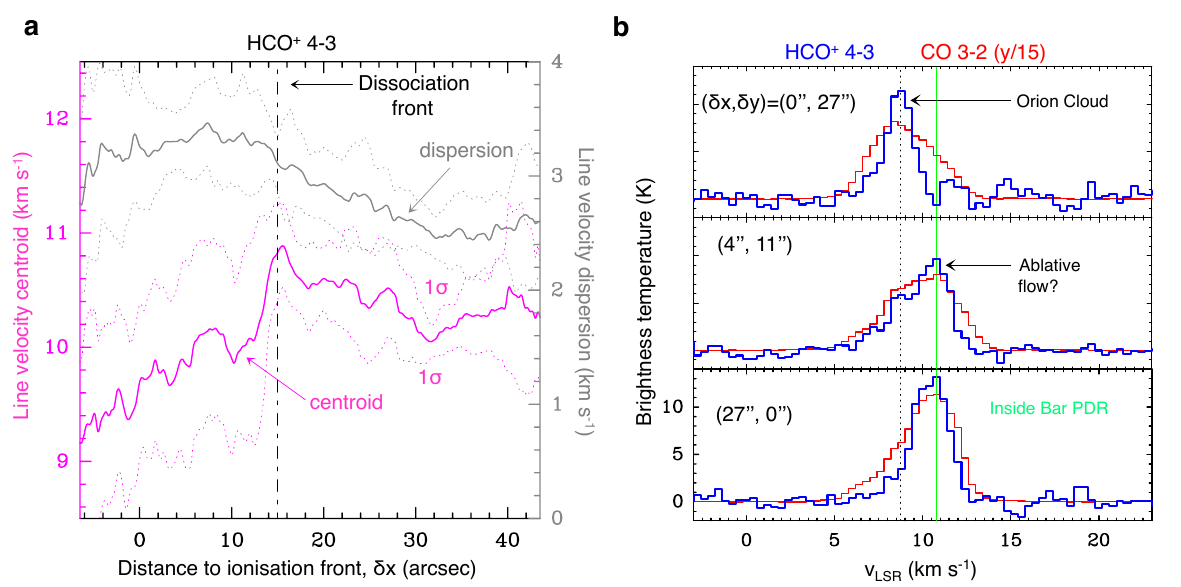}
\caption{\textbf{Line velocity centroid, dispersion and profiles.\\  a,} 
Vertically-averaged cuts perpendicular to the Bar in the HCO$^+$ $J$=4-3 line velocity centroid  (magenta curve) and FWHM velocity dispersion (grey curve).  \textbf{b}, CO and HCO$^+$ spectra at representative positions. The first two panels (from top to bottom) are positions between the ionisation and dissociation fronts, the third one is inside the molecular Bar. Offsets are given with respect to the rotated images in Extended Data 
Fig.~2. The velocity of the background cloud is v$_{\rm LSR}$$\approx$8.5~km\,s$^{-1}$ (black dashed line), whereas the velocity of the Bar is  v$_{\rm LSR}$$\approx$11~km\,s$^{-1}$ (green line).}
\end{figure}

\renewcommand{\figurename}{Extended Data Table}
\setcounter{figure}{0}

\begin{figure}[h]
\includegraphics[width=1\textwidth]{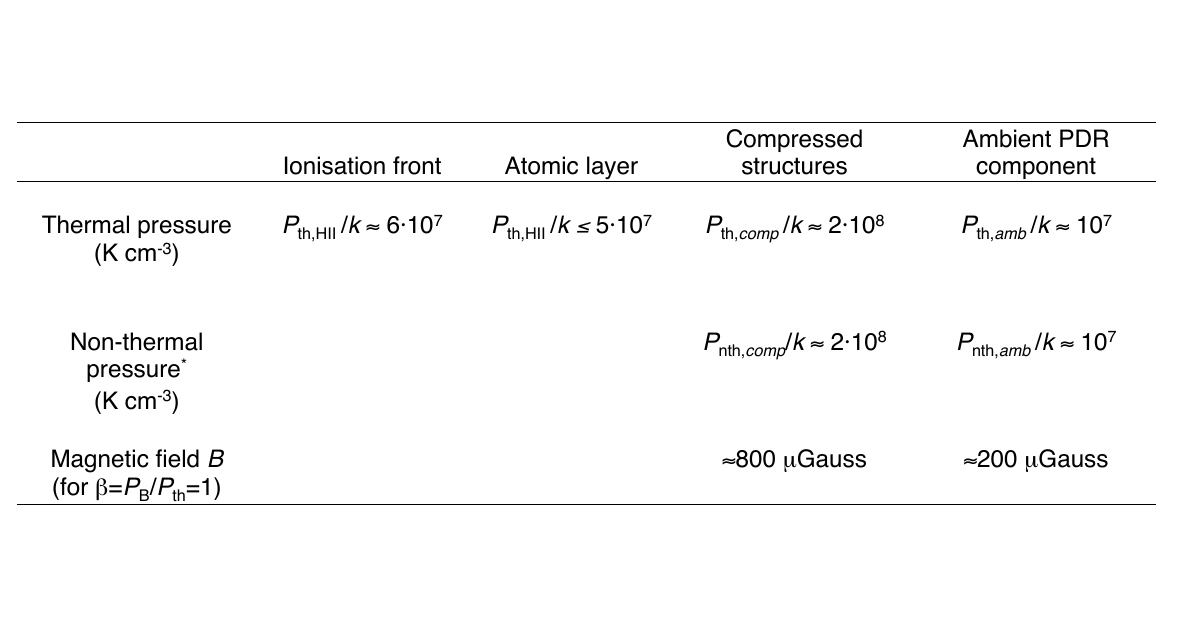}
\caption{\textbf{Gas pressures and estimated magnetic field strengths.}
All values are for a non-thermal velocity dispersion of 
$\sigma_{\rm nth}$$\approx$1~km\,s$^{-1}$.}
\end{figure}

\end{document}